\documentclass[12pt]{article}

\usepackage{epsf}
\usepackage{rotate}

\def\spose#1{\hbox to 0pt{#1\hss}}
\def\lsim{\mathrel{\spose{\lower 3pt\hbox{$\mathchar"218$}}
 \raise 2.0pt\hbox{$\mathchar"13C$}}}
\def\gsim{\mathrel{\spose{\lower 3pt\hbox{$\mathchar"218$}}
 \raise 2.0pt\hbox{$\mathchar"13E$}}}

\begin{document}

\begin{titlepage}

\begin{flushright}
CERN-TH/99-79\\
hep-ph/9903456
\end{flushright}

\vspace{2cm}
\begin{center}
\boldmath
\large\bf New Strategies to Extract $\beta$ and $\gamma$ from\\
\vspace{0.3truecm}
$B_d\to \pi^+\pi^-$ and $B_s\to K^+K^-$
\unboldmath
\end{center}

\vspace{1.2cm}
\begin{center}
Robert Fleischer\\[0.1cm]
{\sl Theory Division, CERN, CH-1211 Geneva 23, Switzerland}
\end{center}

\vspace{1.7cm}
\begin{abstract}
\vspace{0.2cm}\noindent
The decays $B_d\to \pi^+\pi^-$ and $B_s\to K^+K^-$ are related to
each other by interchanging all down and strange quarks, i.e.\ through
the $U$-spin flavour symmetry of strong interactions. A completely 
general parametrization of the CP-violating observables of these modes 
is presented within the framework of the Standard Model, allowing the 
determination of the angles $\beta$ and $\gamma$ of the unitarity triangle. 
This strategy is affected neither by penguin contributions nor by any 
final-state-interaction effects, and its theoretical accuracy is only 
limited by $U$-spin-breaking corrections. If the $B^0_d$--$\overline{B^0_d}$ 
mixing phase $2\beta$ is determined separately, for example with the help 
of $B_d\to J/\psi\, K_{\rm S}$, $\gamma$ can be extracted with a reduced 
$U$-spin flavour symmetry input. A variant of this strategy to determine 
$\gamma$, which uses $B_d\to\pi^\mp K^\pm$ instead of $B_s\to K^+K^-$
and relies -- in addition to the $SU(3)$ flavour symmetry -- on a certain 
dynamical assumption, is also briefly discussed.
\end{abstract}

\vfill
\noindent
CERN-TH/99-79\\
March 1999

\end{titlepage}

\thispagestyle{empty}
\vbox{}
\newpage
 
\setcounter{page}{1}

\section{Introduction}\label{intro}
The exploration of CP violation in the $B$-meson system and the
determination of the three angles $\alpha$, $\beta$ and $\gamma$ 
of the usual non-squashed unitarity triangle \cite{ut} of the 
Cabibbo--Kobayashi--Maskawa matrix (CKM matrix) \cite{ckm} is one
of the central goals of future $B$-physics experiments. However, 
only the determination of $\beta$ with the help of the ``gold-plated'' 
mode $B_d\to J/\psi\, K_{\rm S}$ is ``straightforward'' \cite{bisa}, 
whereas the determination of $\alpha$ and $\gamma$ is considerably
more involved \cite{revs}.

In the literature, the decay $B_d\to\pi^+\pi^-$ usually appears as 
a tool to determine $\alpha=180^\circ-\beta-\gamma$. Unfortunately, 
penguin contributions are expected to affect this determination severely 
\cite{alph}. Although there are several strategies on the market to take 
care of these penguin uncertainties \cite{revs,alph}, they are usually 
very challenging from an experimental point of view. In this paper, a 
new way of making use of the CP-violating observables of the
decay $B_d\to\pi^+\pi^-$ is proposed. Combining them with those of
the mode $B_s\to K^+K^-$, a simultaneous determination of $\beta$
and $\gamma$ becomes possible. The decays $B_d\to\pi^+\pi^-$ and 
$B_s\to K^+K^-$ are related to each other by interchanging all
down and strange quarks, i.e.\ through the so-called ``$U$-spin''
subgroup of the $SU(3)$ flavour symmety of strong interactions. The
utility of $B_s\to K^+K^-$ to probe $\gamma$ was already pointed out
in several previous publications \cite{BsKK}. The new strategy proposed
here is not affected by penguin topologies -- it rather makes use of 
them -- and does not rely on certain ``plausible'' dynamical or 
model-dependent assumptions. Moreover, final-state interaction effects 
\cite{FSI}, which led to considerable attention in the recent literature 
in the context of the determination of $\gamma$ from $B\to\pi K$ decays 
\cite{BpiK}, do not lead to any problems, and the theoretical accuracy is
only limited by $U$-spin-breaking effects. A conceptually quite similar
strategy to extract $\gamma$ using $B_{s(d)}\to J/\psi\, K_{\rm S}$ and 
$B_{d (s)}\to D^{\,+}_{d(s)}\, D^{\,-}_{d(s)}$ decays was recently
proposed in \cite{Uspin-PAP1}. 

The determination of $\gamma$ is a crucial element in the test of the 
Standard Model description of CP violation. In particular, this angle 
should be measured in a variety of ways to check whether one 
finds the same result consistently. During the recent years, 
several methods to accomplish this task were developed \cite{revs}.
Since the $e^+e^-$ $B$-factories operating at the $\Upsilon(4S)$ resonance 
will not be in a position to explore $B_s$ decays, a strong emphasis was
given to decays of non-strange $B$ mesons. However, also the $B_s$ 
system provides interesting strategies to determine $\gamma$ 
(see, for example, \cite{Bsclean}). In order to make use of these methods, 
dedicated $B$-physics experiments at hadron machines, such as LHCb, 
are the natural place. The $B_s$ system exhibits several peculiar features. 
Within the Standard Model, the weak $B^0_s$--$\overline{B^0_s}$ mixing 
phase is very small, and studies of $B_s$ decays involve very rapid 
$B^0_s$--$\overline{B^0_s}$ oscillations, which are due to the large mass 
difference $\Delta M_s\equiv M_{\rm H}^{(s)}-M_{\rm L}^{(s)}$ between the 
mass eigenstates $B_s^{\rm H}$ (``heavy'') and $B_s^{\rm L}$ (``light''). 
Future $B$-physics experiments at hadron machines should be in a position 
to resolve these oscillations. In contrast to the $B_d$ case, there may be 
a sizeable width difference $\Delta\Gamma_s\equiv\Gamma_{\rm H}^{(s)}-
\Gamma_{\rm L}^{(s)}$ between the $B_s$ mass eigenstates \cite{DGamma}, 
which may allow studies of CP violation with ``untagged'' 
$B_s$ data samples, where one does not distinguish between initially, 
i.e.\ at time $t=0$, present $B_s^0$ or $\overline{B^0_s}$ mesons 
\cite{dunietz}. In such untagged rates, the rapid $B^0_s$--$\overline{B^0_s}$ 
oscillations cancel.

There are a few theoretically clean strategies to determine $\gamma$, 
making use of pure ``tree'' decays, for example of $B^\pm\to D K^\pm$ 
or $B_s\to D_s^\pm K^\mp$ modes. Since no flavour-changing neutral-current 
(FCNC) processes contribute to the corresponding decay amplitudes, it is 
quite unlikely that they -- and the extracted value of $\gamma$ -- are 
affected significantly by new physics. In contrast, the strategies
discussed in this paper rely on interference effects between ``tree'' 
and ``penguin'', i.e.\ FCNC, processes. Therefore, new physics may well 
show up in the corresponding decay amplitudes, thereby affecting the 
CP-violating observables and the extracted value of $\gamma$.

The outline of this paper is as follows: in Section~\ref{Obs}, a completely
general parametrization of the $B_d\to\pi^+\pi^-$ and $B_s\to K^+K^-$
decay amplitudes, as well as of the corresponding CP-violating observables, 
is given within the Standard Model. The strategies to determine $\beta$ and 
$\gamma$ with the help of these observables are discussed in 
Section~\ref{Bet-Gam-Det}, where also an approach, which uses 
$B_d\to\pi^\mp K^\pm$ instead of $B_s\to K^+K^-$ and relies -- in addition
to the $SU(3)$ flavour symmetry -- on a certain dynamical assumption, 
is briefly discussed. In Section~\ref{Break}, the $U$-spin-breaking 
corrections affecting these strategies are investigated in more detail, 
and the conclusions are summarized in Section~\ref{concl}.

\begin{figure}
\begin{center}
\leavevmode
\epsfysize=5.5truecm 
\epsffile{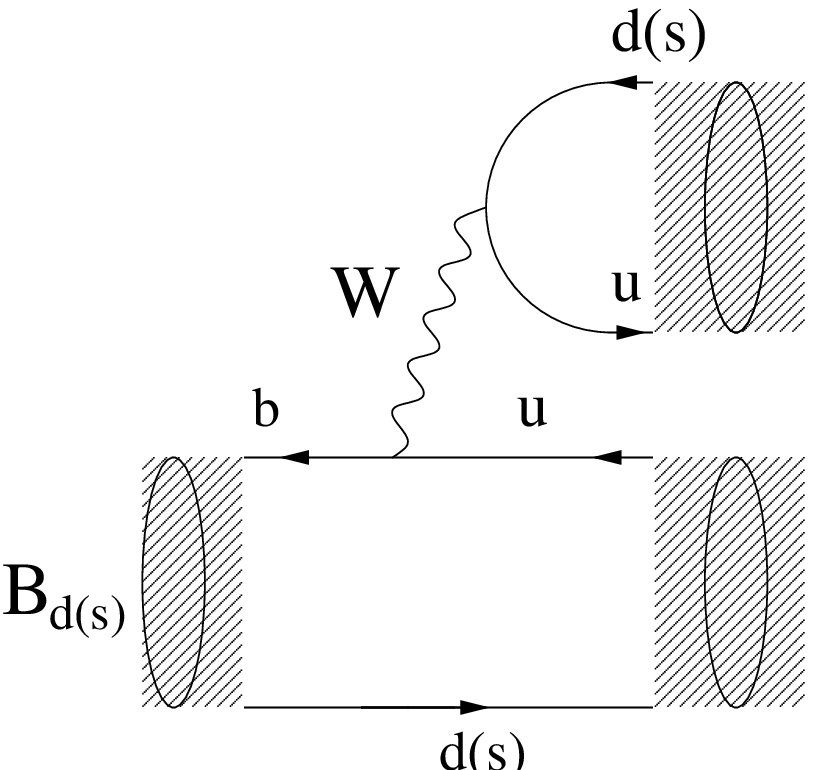} \hspace*{1truecm}
\epsfysize=5.5truecm 
\epsffile{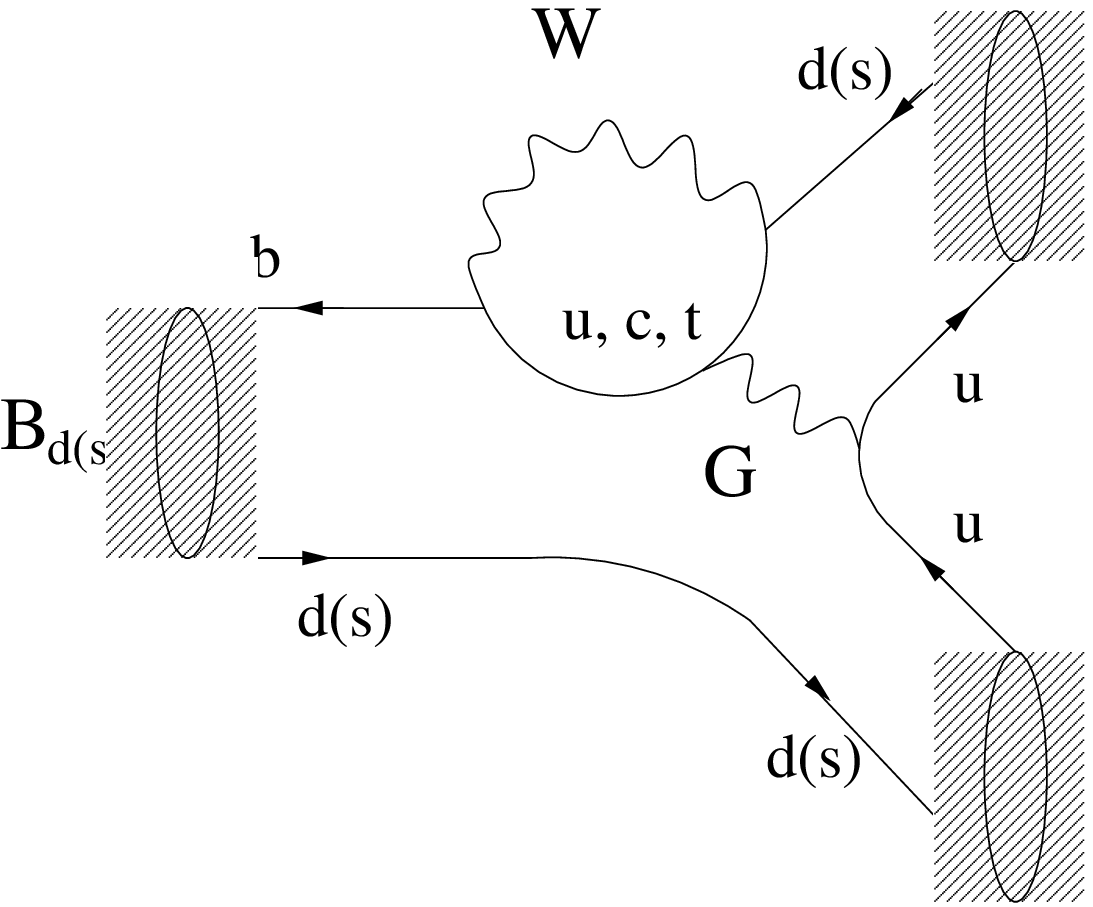}
\end{center}
\caption{Feynman diagrams contributing to 
$B_d\to\pi^+\pi^-$ and $B_s\to K^+K^-$.}\label{fig:BPIPIKK}
\end{figure}

\boldmath
\section{Decay Amplitudes and CP-violating Observables}\label{Obs}
\unboldmath
The decay $B_d^0\to\pi^+\pi^-$ originates from $\bar b\to\bar uu\bar d$ 
quark-level processes, as can be seen in Fig.\ \ref{fig:BPIPIKK}. Its 
transition amplitude can be written as
\begin{equation}\label{Bd-ampl1}
A(B_d^0\to\pi^+\pi^-)=\lambda_u^{(d)}\left(A_{\rm cc}^{u}+A_{\rm pen}^{u}
\right)+\lambda_c^{(d)}A_{\rm pen}^{c}+\lambda_t^{(d)}A_{\rm pen}^{t}\,,
\end{equation}
where $A_{\rm cc}^{u}$ is due to current--current contributions, and the 
amplitudes $A_{\rm pen}^{j}$ describe penguin topologies with internal $j$ 
quarks ($j\in\{u,c,t\})$. These penguin amplitudes take into account both 
QCD and electroweak penguin contributions. The quantities
\begin{equation}
\lambda_j^{(d)}\equiv V_{jd}V_{jb}^\ast
\end{equation}
are the usual CKM factors. If we make use of the unitarity of the CKM matrix
and apply the Wolfenstein parametrization \cite{wolf}, generalized to include 
non-leading terms in $\lambda$ \cite{blo}, we obtain
\begin{equation}\label{Bd-ampl2}
A(B_d^0\to\pi^+\pi^-)=e^{i\gamma}\left(1-\frac{\lambda^2}{2}\right){\cal C}
\left[1-d\,e^{i\theta}e^{-i\gamma}\right],
\end{equation}
where
\begin{equation}\label{Aap-def}
{\cal C}\equiv\lambda^3A\,R_b\left(A_{\rm cc}^{u}+A_{\rm pen}^{ut}\right)
\end{equation}
with $A_{\rm pen}^{ut}\equiv A_{\rm pen}^{u}-A_{\rm pen}^{t}$, and
\begin{equation}\label{ap-def}
d\,e^{i\theta}\equiv\frac{1}{(1-\lambda^2/2)R_b}
\left(\frac{A_{\rm pen}^{ct}}{A_{\rm cc}^{u}+A_{\rm pen}^{ut}}\right).
\end{equation}
The quantity $A_{\rm pen}^{ct}$ is defined in analogy to $A_{\rm pen}^{ut}$,
and the CKM factors are given by
\begin{equation}\label{CKM-exp}
\lambda\equiv|V_{us}|=0.22\,,\quad A\equiv\frac{1}{\lambda^2}
\left|V_{cb}\right|=0.81\pm0.06\,,\quad R_b\equiv\frac{1}{\lambda}
\left|\frac{V_{ub}}{V_{cb}}\right|=0.41\pm0.07\,.
\end{equation}
For the following considerations, time-dependent CP asymmetries play a key 
role. In the case of a general $B_d$ decay into a final CP eigenstate 
$|f\rangle$, satisfying 
\begin{equation}
({\cal CP})|f\rangle=\eta\,|f\rangle, 
\end{equation}
we have~\cite{revs}
\begin{eqnarray}
\lefteqn{a_{\rm CP}(B_d(t)\to f)\equiv
\frac{\Gamma(B_d^0(t)\to f)-
\Gamma(\overline{B_d^0}(t)\to f)}{\Gamma(B_d^0(t)\to f)+
\Gamma(\overline{B_d^0}(t)\to f)}}\nonumber\\
&&={\cal A}_{\rm CP}^{\rm dir}(B_d\to f)\cos(\Delta M_d t)+
{\cal A}_{\rm CP}^{\rm mix}(B_d\to f)\sin(\Delta M_d t)\,.
\end{eqnarray}
If the $B^0_d\to f$ decay amplitude takes the same form as (\ref{Bd-ampl2}), 
we obtain the following expressions for the ``direct'' and ``mixing-induced'' 
CP-violating observables:
\begin{eqnarray}
{\cal A}_{\rm CP}^{\rm dir}(B_d\to f)&=&
-\left[\frac{2\,d\sin\theta\sin\gamma}{1-
2\,d\cos\theta\cos\gamma+d^2}\right]\label{ACP-dir-d}\\
{\cal A}_{\rm CP}^{\rm mix}(B_d\to f)&=&\eta\left[\,
\frac{\sin(\phi_d+2\gamma)-2\,d\,\cos\theta\,\sin(\phi_d+\gamma)+
d^2\sin\phi_d}{1-2\,d\cos\theta\cos\gamma+d^2}\,\right],\label{ACP-mix-d}
\end{eqnarray}
where $\phi_d=2\beta$ denotes the $B^0_d$--$\overline{B^0_d}$ mixing phase,
which can be determined, for instance, with the help of the ``gold-plated'' 
mode $B_d\to J/\psi\,K_{\rm S}$. Strictly speaking, mixing-induced CP
violation in $B_d\to J/\psi\, K_{\rm S}$ probes $2\beta+\phi_K$, where
$\phi_K$ is related to the weak $K^0$--$\overline{K^0}$ mixing phase and 
is negligibly small in the Standard Model. Due to the small value of the
CP-violating parameter $\varepsilon_K$ of the neutral kaon system, 
$\phi_K$ can only be affected by very contrived models of new physics 
\cite{nirsil}.

In the case of $B_d\to\pi^+\pi^-$, $\eta$ is equal to $+1$, and for 
negligible values of the ``penguin parameter'' $d$, we have 
${\cal A}_{\rm CP}^{\rm mix}(B_d\to \pi^+\pi^-)=+\sin(\phi_d+2\gamma)=
\sin(2\beta+2\gamma)=-\sin(2\alpha)$. However, the penguin contributions 
are expected to play an important role. This feature is already indicated 
experimentally by recent CLEO results on $B\to\pi K$ modes.

Let us now turn to the decay $B_s^0\to K^+K^-$. It originates from 
$\bar b\to \bar uu\bar s$ quark-level processes, as can be seen in 
Fig.\ \ref{fig:BPIPIKK}. Using a notation similar to that in 
(\ref{Bd-ampl2}), we obtain
\begin{equation}\label{Bs-ampl}
A(B_s^0\to K^+K^-)=e^{i\gamma}\lambda\,{\cal C}'\left[1+\left(
\frac{1-\lambda^2}{\lambda^2}\right)d'e^{i\theta'}e^{-i\gamma}\right],
\end{equation}
where
\begin{equation}
{\cal C}'\equiv\lambda^3A\,R_b\left(A_{\rm cc}^{u'}+A_{\rm pen}^{ut'}\right)
\end{equation}
and 
\begin{equation}\label{dp-def}
d'e^{i\theta'}\equiv\frac{1}{(1-\lambda^2/2)R_b}
\left(\frac{A_{\rm pen}^{ct'}}{A_{\rm cc}^{u'}+A_{\rm pen}^{ut'}}\right)
\end{equation}
correspond to (\ref{Aap-def}) and (\ref{ap-def}), respectively. The primes
remind us that we are dealing with a $\bar b\to\bar s$ transition. It
should be emphasized that (\ref{Bd-ampl2}) and (\ref{Bs-ampl}) are
completely general parametrizations of the $B_d^0\to\pi^+\pi^-$ and
$B_s^0\to K^+K^-$ decay amplitudes within the Standard Model, relying only 
on the unitarity of the CKM matrix. In particular, these expressions take 
into account also final-state interaction effects, which can be considered 
as long-distance penguin topologies with internal up- and charm-quark 
exchanges \cite{bfm}.

As we have already noted, there may be a sizeable width difference 
$\Delta\Gamma_s\equiv\Gamma_{\rm H}^{(s)}-\Gamma_{\rm L}^{(s)}$ between 
the $B_s$ mass eigenstates \cite{DGamma}, which may allow studies of CP 
violation with ``untagged'' $B_s$ data samples \cite{dunietz}. Such 
untagged rates take the following form:
\begin{equation}\label{untagged}
\Gamma(B^0_s(t)\to f)+\Gamma(\overline{B^0_s}(t)\to f)\propto
R_{\rm H}\,e^{-\Gamma_{\rm H}^{(s)}t}+R_{\rm L}\,e^{-\Gamma_{\rm L}^{(s)}t},
\end{equation}
whereas the time-dependent CP asymmetry is given by
\begin{eqnarray}
\lefteqn{a_{\rm CP}(B_s(t)\to f)\equiv\frac{\Gamma(B^0_s(t)\to f)-
\Gamma(\overline{B^0_s}(t)\to f)}{\Gamma(B^0_s(t)\to f)+
\Gamma(\overline{B^0_s}(t)\to f)}}\nonumber\\
&&=2\,e^{-\Gamma_s t}\left[\frac{{\cal A}_{\rm CP}^{\rm dir}(B_s\to f)
\cos(\Delta M_s t)+{\cal A}_{\rm CP}^{\rm mix}(B_s\to f)\sin(\Delta M_s t)}{
e^{-\Gamma_{\rm H}^{(s)}t}+e^{-\Gamma_{\rm L}^{(s)}t}+
{\cal A}_{\rm \Delta\Gamma}(B_s\to f)\left(e^{-\Gamma_{\rm H}^{(s)}t}-
e^{-\Gamma_{\rm L}^{(s)}t}\right)} \right]
\end{eqnarray}
with ${\cal A}_{\Delta\Gamma}(B_s\to f)=(R_{\rm H}-R_{\rm L})/(R_{\rm H}+
R_{\rm L})$. If the $B_s^0\to f$ decay amplitude takes the same form as 
(\ref{Bs-ampl}), we have
\begin{eqnarray}\label{ACP-s}
{\cal A}_{\rm CP}^{\rm dir}(B_s\to f)&=&
+\left[\frac{2\,\tilde d'\sin\theta'\sin\gamma}{1+
2\,\tilde d'\cos\theta'\cos\gamma+\tilde d'^2}\right]\label{ACPdir-s}\\
{\cal A}_{\rm CP}^{\rm mix}(B_s\to f)&=&+\,\eta\left[\,
\frac{\sin(\phi_s+2\gamma)+2\,\tilde d'\,\cos\theta'\,\sin(\phi_s+\gamma)+
\tilde d'^2\sin\phi_s}{1+2\,\tilde d'\cos\theta'\cos\gamma+\tilde d'^2}\,
\right]\label{ACPmix-s}\\
{\cal A}_{\Delta\Gamma}(B_s\to f)&=&
-\,\eta\left[\frac{\cos(\phi_s+2\gamma)+2\,\tilde d'\cos\theta'
\cos(\phi_s+\gamma)+\tilde d'^2\cos\phi_s}{1+2\,\tilde d'\cos\theta'
\cos\gamma+\tilde d'^2}\right].\label{ADG-s}
\end{eqnarray}
These observables are not independent quantities, and satisfy 
the relation 
\begin{equation}\label{Obs-rel}
\Bigl[{\cal A}_{\rm CP}^{\rm dir}(B_s\to f)\Bigr]^2+
\Bigl[{\cal A}_{\rm CP}^{\rm mix}(B_s\to f)\Bigr]^2+
\Bigl[{\cal A}_{\Delta\Gamma}(B_s\to f)\Bigr]^2=1.
\end{equation}
In the general expressions (\ref{ACPdir-s})--(\ref{ADG-s}), we have 
introduced the abbreviation
\begin{equation}\label{d-tilde}
\tilde d'\equiv\left(\frac{1-\lambda^2}{\lambda^2}\right)d',
\end{equation}
and $\phi_s\equiv-2\delta\gamma=2\,\mbox{arg}(V_{ts}^\ast V_{tb})$ denotes 
the $B^0_s$--$\overline{B^0_s}$ mixing phase. Within the Standard Model,
we have $2\delta\gamma\approx0.03$ due to a Cabibbo suppression of
${\cal O}(\lambda^2)$, implying that $\phi_s$ is very small. This phase 
can be probed -- and even determined -- with the help of the decay 
$B_s\to J/\psi\, \phi$ (see, for example, \cite{ddf1}). Large 
CP-violating effects in this decay would signal that $2\delta\gamma$
is not tiny, and would be a strong indication for new-physics 
contributions to $B^0_s$--$\overline{B^0_s}$ mixing.

\boldmath
\section{Extracting $\beta$ and $\gamma$}\label{Bet-Gam-Det}
\unboldmath
Since the decays $B_d\to\pi^+\pi^-$ and $B_s\to K^+K^-$ are related
to each other by interchanging all strange and down quarks, the $U$-spin
flavour symmetry of strong interactions implies
\begin{eqnarray}
d'&=&d\label{Rel1}\\
\theta'&=&\theta.\label{Rel2}
\end{eqnarray}
In contrast to certain isospin relations, electroweak penguins do not
lead to any problems in the $U$-spin relations (\ref{Rel1}) and (\ref{Rel2}). 
Consequently, if we assume that the $B^0_s$--$\overline{B^0_s}$ mixing phase 
$\phi_s$ is negligibly small, or that it is determined with the help of the
decay $B_s\to J/\psi\, \phi$, the four observables provided by the 
time-dependent CP asymmetries of the modes $B_d\to\pi^+\pi^-$ and 
$B_s\to K^+K^-$ depend on the four ``unknowns'' $d$, $\theta$, 
$\phi_d=2\beta$ and $\gamma$ in the strict $U$-spin limit. These quantities 
can therefore be determined simultaneously. In order to extract $\gamma$, 
it suffices to consider ${\cal A}_{\rm CP}^{\rm mix}(B_s\to K^+K^-)$ and 
the direct CP asymmetries ${\cal A}_{\rm CP}^{\rm dir}(B_s\to K^+K^-)$ and 
${\cal A}_{\rm CP}^{\rm dir}(B_d\to\pi^+\pi^-)$. If we make use, in addition,
of ${\cal A}_{\rm CP}^{\rm mix}(B_d\to\pi^+\pi^-)$, $\beta$ can be determined 
as well. 

Let us now give the formulae to implement this approach in a mathematical
way. Using the general parametrization of the CP-violating $B_s\to K^+K^-$ 
observables given in the previous section, we obtain
\begin{equation}\label{ddet}
\tilde d'=\sqrt{\frac{1}{k'}\left[l'\pm\sqrt{l'^2-h'k'}\right]}
\end{equation}
\begin{equation}\label{dcos}
2\,\tilde d'\cos\theta'=-\left(u'+v'\tilde d'^2\right)
\end{equation}
\begin{equation}\label{dsin}
2\,\tilde d'\sin\theta'=\left[1-u'\cos\gamma+(1-v'\cos\gamma)\,\tilde d'^2
\right]\left[\frac{{\cal A}_{\rm CP}^{\rm dir}(B_s\to K^+K^-)}{\sin
\gamma}\right],
\end{equation}
where
\begin{eqnarray}
h'&=&u'^2+D'(1-u'\cos\gamma)^2\label{h-def}\\
k'&=&v'^2+D'(1-v'\cos\gamma)^2\\
l'&=&2-u'v'-D'(1-u'\cos\gamma)(1-v'\cos\gamma)\label{l-def}
\end{eqnarray}
with 
\begin{eqnarray}
u'&=&\frac{{\cal A}_{\rm CP}^{\rm mix}(B_s\to K^+K^-)-\sin(\phi_s
+2\gamma)}{{\cal A}_{\rm CP}^{\rm mix}(B_s\to K^+K^-)\cos\gamma-
\sin(\phi_s+\gamma)},\label{us-def}\\
v'&=&\frac{{\cal A}_{\rm CP}^{\rm mix}(B_s\to K^+K^-)-
\sin\phi_s}{{\cal A}_{\rm CP}^{\rm mix}(B_s\to K^+K^-)\cos\gamma-
\sin(\phi_s+\gamma)},\label{vs-def}
\end{eqnarray}
and
\begin{equation}
D'=\left[\frac{{\cal A}_{\rm CP}^{\rm dir}(B_s\to K^+K^-)}{\sin
\gamma}\right]^2.
\end{equation}
Taking into account (\ref{d-tilde}) and the $U$-spin relations (\ref{Rel1}) 
and (\ref{Rel2}), these expressions allow us to determine the direct CP 
asymmetry ${\cal A}_{\rm CP}^{\rm dir}(B_d\to\pi^+\pi^-)$ (see 
(\ref{ACP-dir-d})) as a function of $\gamma$. The measured value of 
${\cal A}_{\rm CP}^{\rm dir}(B_d\to\pi^+\pi^-)$ then fixes $\gamma$ and
$d\,e^{i\theta}$. Inserting $\gamma$ and $d\,e^{i\theta}$ thus determined 
into (\ref{ACP-mix-d}), the measured value of the mixing-induced
CP asymmetry ${\cal A}_{\rm CP}^{\rm mix}(B_d\to\pi^+\pi^-)$ allows us
to determine $\beta$. The $U$-spin-breaking corrections to (\ref{Rel1}) 
and (\ref{Rel2}) will be discussed in Section~\ref{Break}. Needless to 
note that also $d\,e^{i\theta}$ is an interesting quantity, allowing valuable 
insights into the hadronization dynamics of $B_d\to\pi^+\pi^-$.

\begin{figure}
\centerline{\rotate[r]{
\epsfysize=11.1truecm
{\epsffile{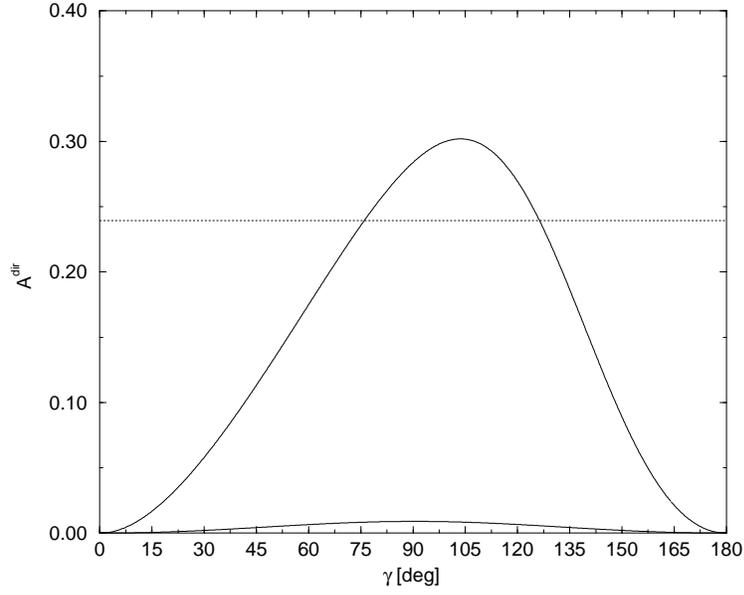}}}}
\caption{The dependence of ${\cal A}_{\rm CP}^{\rm dir}(B_d\to\pi^+\pi^-)$
on $\gamma$ fixed through the CP-violating $B_s\to K^+K^-$ observables 
for a specific example discussed in the text.}\label{fig:ACPdir}
\end{figure}

\begin{figure}
\centerline{\rotate[r]{
\epsfysize=11.1truecm
{\epsffile{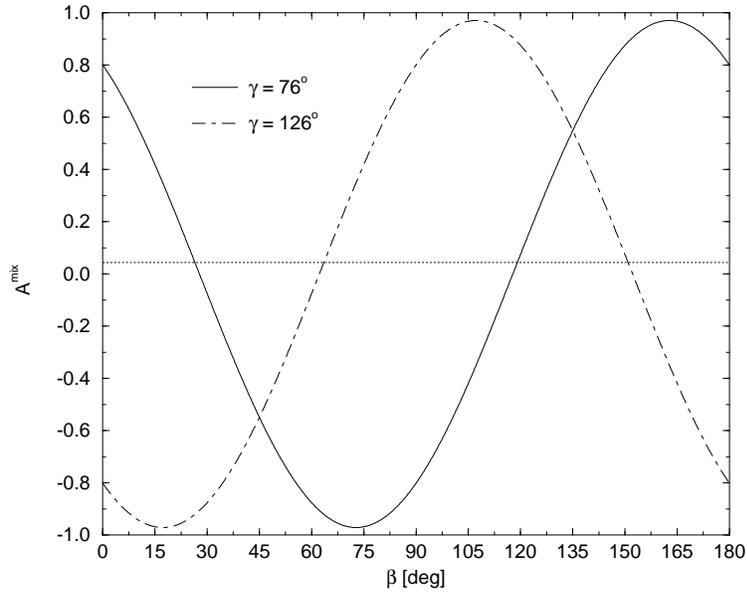}}}}
\caption{The dependence of ${\cal A}_{\rm CP}^{\rm mix}(B_d\to\pi^+\pi^-)$
on $\beta$ fixed through the CP-violating $B_s\to K^+K^-$ observables for 
a specific example discussed in the text.}\label{fig:ACPmix}
\end{figure}

Since the strategy discussed in the previous paragraph appears quite
abstract, let us illustrate it by considering a simple example. 
If we neglect the $B^0_s$--$\overline{B^0_s}$ mixing phase $\phi_s$ and
assume $2\beta=53^\circ$ and $\gamma=76^\circ$, lying within 
the ranges allowed at present for these angles, implied by the usual 
indirect fits of the unitarity triangle, as well as $d=d'=0.3$ and 
$\theta=\theta'=210^\circ$, we obtain 
\begin{equation}
\begin{array}{lcllcl}
{\cal A}_{\rm CP}^{\rm dir}(B_d\to\pi^+\pi^-)&=&+24\%,\quad& 
{\cal A}_{\rm CP}^{\rm mix}(B_d\to\pi^+\pi^-)&=&+4.4\%,\\ 
{\cal A}_{\rm CP}^{\rm dir}(B_s\to K^+K^-)&=&-17\%,\quad&
{\cal A}_{\rm CP}^{\rm mix}(B_s\to K^+K^-)&=&-28\%.
\end{array}
\end{equation}
In Fig.\ \ref{fig:ACPdir}, we show the dependence of 
${\cal A}_{\rm CP}^{\rm dir}(B_d\to\pi^+\pi^-)$ on $\gamma$, obtained as
described above. The dotted line in this figure represents the ``measured''
value of $+24\%$. In this example, we get two solutions for $\gamma$, the
``true'' value of $76^\circ$, and a second one of $126^\circ$. Inserting 
these values for $\gamma$ and the corresponding ones for $d$ and $\theta$ 
into the expression (\ref{ACP-mix-d}) for ${\cal A}_{\rm CP}^{\rm mix}
(B_d\to\pi^+\pi^-)$, we obtain the curves shown in Fig.\ \ref{fig:ACPmix}, 
where the solid line corresponds to $\gamma=76^\circ$, and the dot-dashed 
line corresponds to the second solution of $\gamma=126^\circ$. The dotted 
line represents the ``measured'' value ${\cal A}_{\rm CP}^{\rm mix}
(B_d\to\pi^+\pi^-)=+4.4\%$. If we look at Fig.\ \ref{fig:ACPmix}, we observe
that $\beta$ can be determined in this example up to a fourfold ambiguity. 
Here we have assumed that $\beta\in[0^\circ,180^\circ]$, as implied by
the measured value of $\varepsilon_K$. A similar comment applies to the 
range for $\gamma$. If the twofold ambiguity in the extraction of $\gamma$ 
could be resolved, we were left with a twofold ambiguity for $\beta$.

The $B^0_d$--$\overline{B^0_d}$ mixing phase $\phi_d=2\beta$ can
be determined in a reliable way with the help of the ``gold-plated'' 
mode $B_d\to J/\psi\,K_{\rm S}$ \cite{bisa}. Strictly speaking, a 
measurement of mixing-induced CP violation in $B_d\to J/\psi\,K_{\rm S}$ 
allows us to determine only $\sin(2\beta)$, i.e.\ to fix $\phi_d=2\beta$ 
up to a twofold ambiguity for $\beta\in[0^\circ,180^\circ]$. However, 
several strategies to resolve this ambiguity were proposed in the literature 
\cite{ambig}, which should be feasible for ``second-generation'' $B$-physics 
experiments. Consequently, $\phi_d=2\beta$ should be known reliably and
unambiguously in the era of these experiments, thereby providing a 
different strategy to combine the observables (\ref{ACP-dir-d}) and 
(\ref{ACP-mix-d}) with (\ref{ACPdir-s}) and (\ref{ACPmix-s}). The point 
is that these quantities allow us to fix contours in the $\gamma$--$d$ and 
$\gamma$--$d'$ planes as functions of the $B^0_d$--$\overline{B^0_d}$ and 
$B^0_s$--$\overline{B^0_s}$ mixing phases in a {\it theoretically clean} 
way. In the $B_s\to K^+K^-$ case, these contours are described by
(\ref{ddet}) with (\ref{d-tilde}). On the other hand, in the case of 
$B_d\to\pi^+\pi^-$, we have 
\begin{equation}\label{Ddet}
d=\sqrt{\frac{1}{k}\left[l\pm\sqrt{l^2-h\,k}\right]}
\end{equation}
\begin{equation}\label{Dcos}
2\,d\cos\theta=u+v\,d^2
\end{equation}
\begin{equation}\label{Dsin}
2\,d\sin\theta=-\left[1-u\cos\gamma+(1-v\cos\gamma)\,d^2
\right]\left[\frac{{\cal A}_{\rm CP}^{\rm dir}(B_d\to\pi^+\pi^-)}{\sin
\gamma}\right],
\end{equation}
where $h$, $k$ and $l$ take the same form as (\ref{h-def})--(\ref{l-def})
with $u'$, $v'$ and $D'$ replaced by 
\begin{eqnarray}
u&=&\frac{{\cal A}_{\rm CP}^{\rm mix}(B_d\to\pi^+\pi^-)-\sin(\phi_d
+2\gamma)}{{\cal A}_{\rm CP}^{\rm mix}(B_d\to\pi^+\pi^-)\cos\gamma-
\sin(\phi_d+\gamma)}\label{u-def}\\
v&=&\frac{{\cal A}_{\rm CP}^{\rm mix}(B_d\to\pi^+\pi^-)-
\sin\phi_d}{{\cal A}_{\rm CP}^{\rm mix}(B_d\to\pi^+\pi^-)\cos\gamma-
\sin(\phi_d+\gamma)}\label{v-def}
\end{eqnarray}
and
\begin{equation}
D=\left[\frac{{\cal A}_{\rm CP}^{\rm dir}(B_d\to\pi^+\pi^-)}{\sin
\gamma}\right]^2.
\end{equation}
The contours in the $\gamma$--$d$ plane are fixed through (\ref{Ddet}).
Using the $U$-spin relation (\ref{Rel1}), the intersection of the 
{\it theoretically clean} $B_s\to K^+K^-$ and $B_d\to\pi^+\pi^-$ contours 
described by (\ref{ddet}) (with (\ref{d-tilde})) and (\ref{Ddet}) allows 
us to determine $\gamma$. Analogously, it is possible to fix 
{\it theoretically clean} contours in the $\gamma$--$\theta^{(')}$ planes 
through (\ref{ddet})--(\ref{dsin}) and (\ref{Ddet})--(\ref{Dsin}), and 
to determine $\gamma$ with the help of the $U$-spin relation (\ref{Rel2}).

\begin{figure}
\centerline{\rotate[r]{
\epsfysize=11.1truecm
{\epsffile{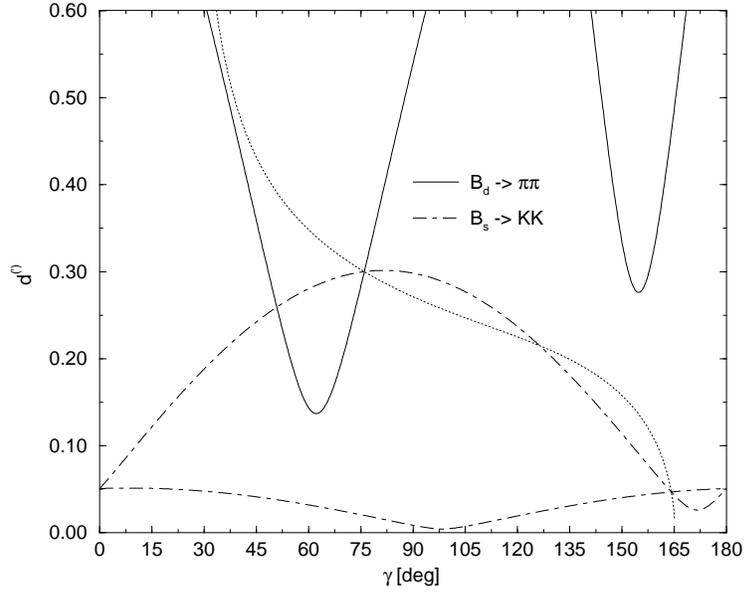}}}}
\caption{The contours in the $\gamma$--$d^{(')}$ planes fixed through the
CP-violating $B_d\to\pi^+\pi^-$ and $B_s\to K^+K^-$ observables for a 
specific example discussed in the text.}\label{fig:cont}
\end{figure}

\begin{figure}
\centerline{\rotate[r]{
\epsfysize=11.1truecm
{\epsffile{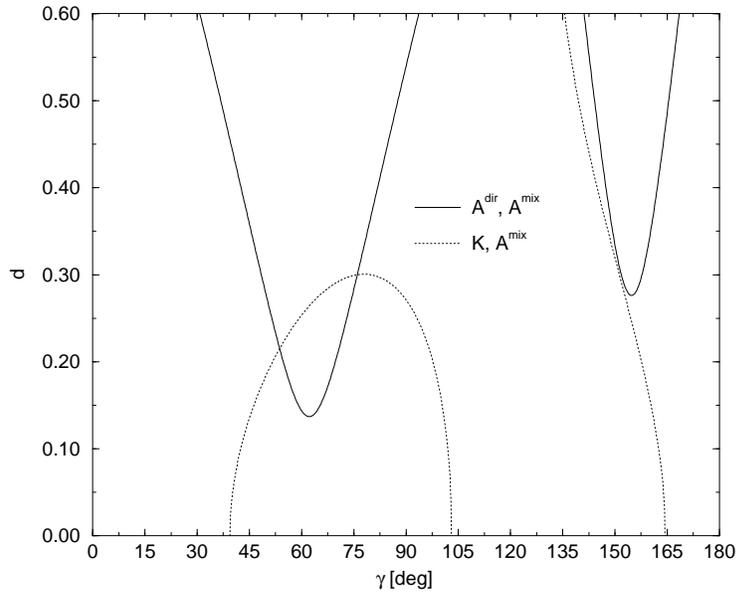}}}}
\caption{The contours in the $\gamma$--$d$ plane fixed through the 
observables ${\cal A}_{\rm CP}^{\rm dir}(B_d\to\pi^+\pi^-)$,
${\cal A}_{\rm CP}^{\rm mix}(B_d\to\pi^+\pi^-)$ and $K$ for a specific
example discussed in the text.}\label{fig:Bdcont}
\end{figure}

In the case of the illustrative example discussed above, we obtain the
contours in the $\gamma$--$d^{(')}$ planes shown in Fig.\ \ref{fig:cont}. 
Here the dot-dashed and solid lines correspond to (\ref{ddet}) and
(\ref{Ddet}), respectively. The intersection of these lines yields a 
twofold solution for $\gamma$, given by $51^\circ$ and 
by the ``true'' value of $76^\circ$. The dotted line in  Fig.\ \ref{fig:cont}
is related to 
\begin{equation}\label{H-intro}
K\equiv-\,\frac{1}{\epsilon}\,
\frac{{\cal A}_{\rm CP}^{\rm dir}(B_d\to\pi^+\pi^-)}{{\cal A}_{\rm 
CP}^{\rm dir}(B_s\to K^+K^-)}=\left(\frac{d\,\sin\theta}{d'\sin\theta'}
\right)\left(\frac{1+2\,\tilde d'\cos\theta'\cos\gamma+
\tilde d'^2}{1-2\,d\cos\theta\cos\gamma+d^2}\right),
\end{equation}
where
\begin{equation}
\epsilon\equiv\frac{\lambda^2}{1-\lambda^2}.
\end{equation}
If we use the $U$-spin relations (\ref{Rel1}) and (\ref{Rel2}), as well
as the mixing-induced CP asymmetry ${\cal A}_{\rm CP}^{\rm mix}
(B_s\to K^+K^-)$, we obtain
\begin{equation}\label{d-H-cont}
d=\epsilon\,\sqrt{\frac{K-1+(1+\epsilon\,K)\,u'\cos\gamma}{1-\epsilon^2K-
(1+\epsilon\,K)\,v'\cos\gamma}}\,,
\end{equation}
which fixes the dotted line in Fig.\ \ref{fig:cont}. Combining all
contours in this figure, we obtain a single solution for $\gamma$ in 
this example, which is given by the ``true'' value of $76^\circ$. 

It is interesting to note that an alternative way to fix certain 
contours through the CP-violating observables 
${\cal A}_{\rm CP}^{\rm dir}(B_d\to\pi^+\pi^-)$ and
${\cal A}_{\rm CP}^{\rm mix}(B_d\to\pi^+\pi^-)$ was proposed in 
\cite{charles}. In this paper, a different parametrization of the
$B_d^0\to\pi^+\pi^-$ decay amplitude was chosen, and in the corresponding
contours, $\alpha$ was set in correlation with a hadronic parameter $|P/T|$,
which represents -- sloppily speaking -- the ratio of the ``penguin'' to the
``tree'' contributions to $B_d^0\to\pi^+\pi^-$. Unfortunately, it is very 
difficult to fix this parameter in a theoretically reliable way, in 
particular as it is also affected by final-state-interaction effects. 
In the strategies proposed above, there are no problems of this kind,
and the theoretical accuracy is only limited by $U$-spin-breaking 
corrections, which will be discussed in more detail in Section~\ref{Break}.

Before turning to these corrections, let us discuss some other interesting
implications of the $U$-spin flavour symmetry. Employing once more 
(\ref{Rel1}) and (\ref{Rel2}), we obtain
\begin{equation}\label{H-det}
K=\frac{1}{\epsilon}\,\left|\frac{{\cal C}}{{\cal C}'}\right|^2
\left[\frac{M_{B_s}}{M_{B_d}}\,\frac{\Phi(M_\pi/M_{B_d},M_\pi/M_{B_d})}{
\Phi(M_K/M_{B_s},M_K/M_{B_s})}\,\frac{\tau_{B_d}}{\tau_{B_s}}\right]
\frac{\mbox{BR}(B_s\to K^+K^-)}{\mbox{BR}(B_d\to\pi^+\pi^-)},
\end{equation}
where
\begin{equation}
\Phi(x,y)\equiv\sqrt{\left[1-(x+y)^2\right]\left[1-(x-y)^2\right]}
\end{equation}
is the usual two-body phase-space function, and BR$(B_s\to K^+K^-)$
and BR$(B_d\to\pi^+\pi^-)$ denote the ``CP-averaged'' branching ratios,
which can be extracted from the ``untagged'' rates (\ref{untagged}).
Moreover, taking into account (\ref{H-intro}), we arrive at the $U$-spin
relation 
\begin{equation}\label{ACP-rel}
\frac{{\cal A}_{\rm CP}^{\rm dir}(B_s\to K^+K^-)}{{\cal A}_{\rm CP}^{\rm dir}
(B_d\to\pi^+\pi^-)}=-\,\left|\frac{{\cal C}'}{{\cal C}}\right|^2
\left[\frac{M_{B_d}}{M_{B_s}}\,\frac{\Phi(M_K/M_{B_s},M_K/M_{B_s})}{
\Phi(M_\pi/M_{B_d},M_\pi/M_{B_d})}\,\frac{\tau_{B_s}}{\tau_{B_d}}\right]
\frac{\mbox{BR}(B_d\to\pi^+\pi^-)}{\mbox{BR}(B_s\to K^+K^-)}.
\end{equation}
An analogous relation holds between the CP-violating asymmetries of the
decays $B^\pm\to \pi^\pm K$ and $B^\pm\to K^\pm K$ \cite{rf}. In the strict
$U$-spin limit, we have $|{\cal C}'|=|{\cal C}|$. Corrections to this
relation can be calculated within the ``factorization'' approximation,
yielding
\begin{equation}\label{U-fact}
\left|\frac{{\cal C}'}{{\cal C}}\right|_{\rm fact}=\,
\frac{f_K}{f_\pi}\frac{F_{B_sK}(M_K^2;0^+)}{F_{B_d\pi}(M_\pi^2;0^+)}
\left(\frac{M_{B_s}^2-M_K^2}{M_{B_d}^2-M_\pi^2}\right),
\end{equation}
where $f_K$ and $f_\pi$ denote the kaon and pion decay constants, and the 
form factors $F_{B_sK}(M_K^2;0^+)$ and $F_{B_d\pi}(M_\pi^2;0^+)$ 
parametrize the hadronic quark-current matrix elements 
$\langle K^-|(\bar b u)_{\rm V-A}|B^0_s\rangle$ and 
$\langle\pi^-|(\bar b u)_{\rm V-A}|B^0_d\rangle$, respectively \cite{BSW}.
Consequently, we have
\begin{equation}\label{Approx-Rel}
\frac{{\cal A}_{\rm CP}^{\rm dir}(B_s\to K^+K^-)}{{\cal A}_{\rm CP}^{\rm dir}
(B_d\to\pi^+\pi^-)}\approx-\,1.56\times
\left[\frac{F_{B_sK}(M_K^2;0^+)}{F_{B_d\pi}(M_\pi^2;0^+)}\right]^2
\times\frac{\tau_{B_s}}{\tau_{B_d}}\times
\frac{\mbox{BR}(B_d\to\pi^+\pi^-)}{\mbox{BR}(B_s\to K^+K^-)}\,,
\end{equation}
where $F_{B_sK}(M_K^2;0^+)\approx F_{B_d\pi}(M_\pi^2;0^+)$ and
$\tau_{B_s}\approx\tau_{B_d}$. As we will see in the following section,
the $U$-spin relations (\ref{Rel1}) and (\ref{Rel2}) do {\it not} receive 
$U$-spin-breaking corrections within the ``factorization'' approximation. 
However, it should be emphasized that also non-factorizable contributions, 
which are not included in (\ref{U-fact}) and in the analysis of 
Section~\ref{Break}, may play an important role. 

Interestingly, the ratio $|{\cal C}'/{\cal C}|$ can be determined with the 
help of (\ref{ACP-rel}), allowing valuable experimental insights into 
$U$-spin breaking. Alternatively, (\ref{H-det}) and (\ref{U-fact}) allow us 
to determine $K$ by using only the CP-averaged $B_d\to\pi^+\pi^-$ and 
$B_s\to K^+K^-$ branching ratios. Employing once again the $U$-spin 
relations (\ref{Rel1}) and (\ref{Rel2}), as well as the general expression 
(\ref{ACP-mix-d}) for the mixing-induced CP asymmetry 
${\cal A}_{\rm CP}^{\rm mix}(B_d\to\pi^+\pi^-)$, we obtain
\begin{equation}\label{H-cont}
d=\sqrt{\frac{\epsilon^2\,(K-1)-\epsilon\,(1+\epsilon\,K)\,u\cos\gamma}{1-
\epsilon^2K+\epsilon\,(1+\epsilon\,K)\,v\cos\gamma}}\,,
\end{equation}
which corresponds to (\ref{d-H-cont}). Consequently, if $\phi_d=2\beta$ is 
determined through $B_d\to J/\psi\,K_{\rm S}$, the CP-violating observables 
${\cal A}_{\rm CP}^{\rm dir}(B_d\to\pi^+\pi^-)$,  
${\cal A}_{\rm CP}^{\rm mix}(B_d\to\pi^+\pi^-)$
and the CP-averaged $B_d\to\pi^+\pi^-$, $B_s\to K^+K^-$ branching ratios
allow us to determine $\gamma$ with the help of the contours in the
$\gamma$--$d$ plane described by (\ref{Ddet}) and (\ref{H-cont}). This
approach is illustrated in Fig.~\ref{fig:Bdcont}, where the solid and
dotted lines correspond to (\ref{Ddet}) and (\ref{H-cont}), respectively.
Although this approach suffers from theoretical uncertainties due to 
(\ref{U-fact}) larger than those of the strategies described above, it does 
not require a time-dependent measurement of $B_s\to K^+K^-$, i.e.\ to 
resolve the $\Delta M_st$ oscillations in this decay. 

Since the decays $B_s\to K^+K^-$ and $B_d\to\pi^\mp K^\pm$ differ only 
in their spectator quarks, we have
\begin{equation}\label{ACP-rep}
{\cal A}_{\rm CP}^{\rm dir}(B_s\to K^+K^-)\approx{\cal A}_{\rm CP}^{\rm dir}
(B_d\to\pi^\mp K^\pm)
\end{equation}
\begin{equation}\label{BR-rep}
\mbox{BR}(B_s\to K^+K^-)
\approx\mbox{BR}(B_d\to\pi^\mp K^\pm)\,\frac{\tau_{B_s}}{\tau_{B_d}},
\end{equation}
allowing us to fix $K$ through the $B_d\to\pi^\mp K^\pm$ and 
$B_d\to\pi^+\pi^-$ observables with the help of the formulae given above. 
In contrast to $B_d\to\pi^\mp K^\pm$, $B_s\to K^+K^-$ receives also 
contributions from ``exchange'' and ``penguin annihilation'' topologies, 
which are usually expected to play a minor role \cite{ghlr}, but may be 
enhanced through certain rescattering effects \cite{FSI}. Although these 
topologies do {\it not} lead to any problems for the strategies using 
$B_d\to\pi^+\pi^-$ and $B_s\to K^+K^-$ discussed in this section -- even 
if they should turn out to be sizeable -- they may affect (\ref{ACP-rep})
and (\ref{BR-rep}). Consequently, these relations rely -- in addition to 
the $SU(3)$ flavour symmetry -- on a certain dynamical assumption. If we 
perform the appropriate replacements in (\ref{ACP-rel}) and (\ref{U-fact}), 
we obtain
\begin{equation}\label{ACP-rel2}
\frac{{\cal A}_{\rm CP}^{\rm dir}(B_d\to\pi^\mp K^\pm)}{{\cal A}_{\rm 
CP}^{\rm dir}(B_d\to\pi^+\pi^-)}\approx-\left(\frac{f_K}{f_\pi}\right)^2
\left[\frac{\Phi(M_\pi/M_{B_d},M_K/M_{B_d})}{\Phi(M_\pi/M_{B_d},M_\pi/M_{B_d})}
\right]\frac{\mbox{BR}(B_d\to\pi^+\pi^-)}{\mbox{BR}(B_d\to\pi^\mp K^\pm)}\,,
\end{equation}
which provides an interesting cross check. The importance of the ``exchange'' 
and ``penguin annihilation'' topologies contributing to $B_s\to K^+K^-$
can be probed with the help of the decay $B_s\to\pi^+\pi^-$. The na\"\i ve 
expectation for the corresponding branching ratio is ${\cal O}(10^{-8})$; 
a significant enhancement would signal that the ``exchange'' and ``penguin 
annihilation'' topologies cannot be neglected. Another interesting decay in 
this respect is $B_d\to K^+K^-$, for which already stronger experimental 
constraints exist \cite{groro}.

If the $B_d\to\pi^+\pi^-$ and $B_d\to\pi^\mp K^\pm$ observables are
measured and $\phi_d$ is fixed through $B_d\to J/\psi\,K_{\rm S}$, 
$\gamma$ can be determined with the help of the contours described
by (\ref{Ddet}) and (\ref{H-cont}), as illustrated in 
Fig.~\ref{fig:Bdcont}. All time-dependent measurements that are required 
for this strategy can in principle be performed at the asymmetric 
$e^+e^-$ $B$-factories operating at the $\Upsilon(4S)$ resonance, 
i.e.\ at BaBar or BELLE. Whereas $B_d\to\pi^\mp K^\pm$ has 
already been observed by the CLEO collaboration with a CP-averaged 
branching ratio of $(1.4\pm0.3\pm0.2)\times10^{-5}$, at present only 
the upper limit BR$(B_d\to\pi^+\pi^-)<0.84\times10^{-5}$ (90\% C.L.) 
is available from CLEO \cite{stone}. If we use (\ref{ACP-rel2}) and
$\mbox{BR}(B_s\to K^+K^-)\approx\mbox{BR}(B_d\to\pi^\mp K^\pm)\approx
1.4\times10^{-5}$, we obtain, for the example discussed above,  
$\mbox{BR}(B_d\to\pi^+\pi^-)\approx0.66\times10^{-5}$, satisfying the 
present upper limit from CLEO. After these remarks, let us come back to 
the $B_d\to\pi^+\pi^-$, $B_s\to K^+K^-$ strategies illustrated in Figs.\ 
\ref{fig:ACPdir}--\ref{fig:cont}, and let us investigate the 
impact of the $U$-spin-breaking corrections in the following section.

\boldmath
\section{A Closer Look at the $U$-spin-breaking Corrections}\label{Break}
\unboldmath
In order to analyse non-leptonic $B$-meson decays theoretically, one
uses low-energy effective Hamiltonians, which are calculated in
renormalization-group-improved perturbation theory, and take the following 
form \cite{BBL-rev}:
\begin{equation}\label{heff}
{\cal H}_{{\rm eff}} = \frac{G_{\rm F}}{\sqrt{2}}\left[ 
\lambda_u^{(q)}\sum_{k=1}^2C_k(\mu) Q_k^{uq} +\lambda_c^{(q)}
\sum_{k=1}^2C_k(\mu) Q_k^{cq} -\lambda_t^{(q)} \sum^{10}_{k=3} 
C_k(\mu) Q_k^q \right].
\end{equation}
Here $Q_1^{jq}$ and $Q_2^{jq}$ ($j\in\{u,c\}$, $q\in\{d,s\}$) are the usual 
current--current operators, $Q_3^q$,\ldots,$Q_6^q$ and $Q_7^q$,\ldots,
$Q_{10}^q$ denote the QCD and electroweak penguin operators, respectively, 
and $\mu={\cal O}(m_b)$ is a renormalization scale. Applying the 
Bander--Silverman--Soni mechanism \cite{bss}, and following the formalism 
developed in \cite{pen-calc}, we obtain
\begin{equation}\label{d-approx}
d\,e^{i\theta}\approx\frac{1}{(1-\lambda^2/2)R_b}\left[\frac{{\cal A}_t+
{\cal A}_c}{{\cal A}_{\rm T}+{\cal A}_t+{\cal A}_u}\right],
\end{equation}
where
\begin{eqnarray}
{\cal A}_{\rm T}&=&\frac{1}{3}\,\overline{C}_1+\overline{C}_2\label{AT}\\
{\cal A}_t&=&\frac{1}{3}\left[\overline{C}_3+\overline{C}_9+
\chi\left(\overline{C}_5+\overline{C}_7\right)\right]+\overline{C}_4+
\overline{C}_{10}+\chi\left(\overline{C}_6+\overline{C}_8\right)\label{A0}\\
{\cal A}_j&=&\frac{\alpha_s}{9\pi}\left[\frac{10}{9}-G(m_j,k,m_b)\right]
\left[\overline{C}_2+\frac{1}{3}\frac{\alpha}{\alpha_s}\left(3\,\overline{C}_1
+\overline{C}_2\right)\right]\left(1+\chi\right),\label{Aq}
\end{eqnarray}
with $j\in\{u,c\}$. The coefficients $\overline{C}_k$ refer to $\mu=m_b$ 
and denote the next-to-leading order scheme-independent Wilson coefficient 
functions introduced in \cite{Buras-NLO}. The quantity
\begin{equation}
\chi=\frac{2M_\pi^2}{(m_u+m_d)(m_b-m_u)}
\end{equation}
is due to the use of the equations of motion for the quark fields, whereas 
the function $G(m_j,k,m_b)$ is related to the one-loop penguin matrix 
elements of the current--current operators $Q_{1,2}^{jq}$ with internal 
$j$ quarks. It is given by 
\begin{equation}
G(m_j,k,m_b)=-\,4\int\limits_{0}^{1}\mbox{d}x\,x\,(1-x)\ln\left[\frac{m_j^{2}-
k^{2}\,x\,(1-x)}
{m_b^{2}}\right],
\end{equation}
where $m_j$ is the $j$-quark mass and $k$ denotes some average four-momentum 
of the virtual gluons and photons appearing in corresponding penguin diagrams 
\cite{pen-calc}.  Kinematical considerations at the quark level imply 
the following ``physical'' range for this parameter:
\begin{equation}
\frac{1}{4}\,\mbox{{\scriptsize $\stackrel{<}{\sim}$}}\,
 \frac{k^2}{m_b^2}\,\mbox{{\scriptsize $\stackrel{<}{\sim}$}}\,\frac{1}{2}\,.
\end{equation}
Since the quantity $d\,e^{i\theta}$ is defined in (\ref{ap-def}) as a ratio 
of certain amplitudes, the decay constants and form factors arising typically 
in the ``factorization'' approximation, as can be seen, for example, 
in (\ref{U-fact}), cancel in (\ref{d-approx}). The expression for the 
$B_s\to K^+K^-$ parameter $d'e^{i\theta'}$ takes the same form as 
(\ref{d-approx}), where $\chi$ is replaced in (\ref{A0}) and (\ref{Aq}) by
\begin{equation}
\chi'=\frac{2M_K^2}{(m_u+m_s)(m_b-m_u)}.
\end{equation}
Consequently, in our approach to evaluate $d\,e^{i\theta}$ and 
$d'e^{i\theta'}$, the $U$-spin-breaking corrections are only due to the 
parameters $\chi$ and $\chi'$. However, up to small electromagnetic 
corrections, the chiral structure of strong interactions implies
\begin{equation}\label{mass-rel}
\frac{M_\pi^2}{m_u+m_d}=\frac{M_K^2}{m_u+m_s},
\end{equation}
leading -- among other things -- to the Gell-Mann--Okubo relation 
(see, for example, \cite{georgi}). In our case, this expression has 
the interesting implication 
\begin{equation}
\chi=\chi', 
\end{equation}
so that the $U$-spin relation 
\begin{equation}\label{Uspin-rel}
d\,e^{i\theta}=d'e^{i\theta'}
\end{equation}
is not affected by $U$-spin-breaking corrections within our formalism. 
Moreover, $\chi$ and $\chi'$ are suppressed by the bottom-quark mass.
Although (\ref{d-approx}) is a simplified expression, which may be affected 
by non-factorizable contributions, it strengthens our confidence 
into~(\ref{Uspin-rel}). Unless such non-factorizable effects have a dramatic
impact, the $U$-spin-breaking corrections to this relation are probably
moderate.

Before giving the conclusions in the following section, let us emphasize 
again that the strategies illustrated in Figs.\ \ref{fig:ACPdir} and 
\ref{fig:ACPmix} rely on (\ref{Uspin-rel}), i.e.\ both on (\ref{Rel1}) 
and (\ref{Rel2}), whereas the extraction of $\gamma$ illustrated in 
Fig.\ \ref{fig:cont} makes only use of (\ref{Rel2}). In particular, the 
solid and dot-dashed contours in this figure are {\it theoretically clean}. 
Using (\ref{d-approx}), we obtain values for $d$ and $\theta$ of the 
same order of magnitude as those employed in the example given in the
previous section.

\boldmath
\section{Conclusions}\label{concl}
\unboldmath
The time evolutions of the decays $B_d\to\pi^+\pi^-$ and $B_s\to K^+K^-$ 
provide interesting strategies to extract the angles $\beta$ and $\gamma$ 
of the unitarity triangle of the CKM matrix. These methods, which make use 
of ``penguin'' topologies and are therefore not affected by their presence, 
take into account final-state interaction effects ``automatically''. 
Moreover, they do not rely on any model-dependent or ``plausible'' dynamical 
assumptions, and their theoretical accuracy is only limited by 
$U$-spin-breaking corrections. Within a certain model-dependent approach 
making use -- among other things -- of ``factorization'' to estimate the 
relevant hadronic matrix elements, these $U$-spin-breaking corrections 
vanish. Although this approach is very simplified and may be affected by 
non-factorizable effects, it strengthens our confidence into the $U$-spin 
relations used for the extraction of $\beta$ and $\gamma$ from the decays 
$B_d\to\pi^+\pi^-$ and $B_s\to K^+K^-$. Moreover, an interesting relation 
between the CP-averaged branching ratios of these modes and the corresponding 
direct CP asymmetries may provide valuable experimental insights into 
$U$-spin breaking.

If the $B^0_d$--$\overline{B^0_d}$ mixing phase $\phi_d=2\beta$ is fixed 
through $B_d\to J/\psi\,K_{\rm S}$, certain theoretically clean contours 
can be determined with the help of the CP-violating $B_d\to\pi^+\pi^-$ 
and $B_s\to K^+K^-$ observables, allowing the extraction of $\gamma$ with
a reduced $U$-spin flavour symmetry input. Since both $\bar b\to \bar d$ 
and $\bar b\to\bar s$ ``penguin'', i.e.\ FCNC, processes contribute to the 
$B_d\to\pi^+\pi^-$ and $B_s\to K^+K^-$ decay amplitudes, the extracted value 
of $\gamma$ may well be affected by new physics. In such a case, 
discrepancies would show up with the values of $\gamma$ determined from 
pure ``tree'' decays, for example from $B_s\to D_s^\pm K^\mp$ modes. 
Possible new-physics contributions to $B^0_s$--$\overline{B^0_s}$ mixing 
can be probed through $B_s\to J/\psi\,\phi$, and can also be taken into 
account with the help of this decay. A similar comment applies to the
$B^0_d$--$\overline{B^0_d}$ mixing phase $\phi_d$. If new physics should
shift $\phi_d$ from its Standard Model value, it could still be fixed
through the ``gold-plated'' mode $B_d\to J/\psi\,K_{\rm S}$.

The strategies proposed in this paper are very interesting for 
``second-generation'' $B$-physics experiments performed at hadron machines, 
for example LHCb, where the very interesting physics potential of the $B_s$ 
system can be fully exploited. At the asymmetric $e^+e^-$ $B$-factories 
operating at the $\Upsilon(4S)$ resonance, which will start taking data 
very soon, this is unfortunately not possible. However, there is also 
a variant of the strategy to determine $\gamma$, where $B_d\to\pi^\mp K^\pm$ 
is used instead of $B_s\to K^+K^-$. This approach has the advantage that 
all required time-dependent measurements can in principle be performed 
at the asymmetric $e^+e^-$ machines. On the other hand, it relies -- 
in addition to the $SU(3)$ flavour symmetry -- on the smallness
of certain ``exchange'' and ``penguin annihilation'' topologies, which
may be enhanced by final-state-interaction effects. Consequently, its 
theoretical accuracy cannot compete with the ``second-generation'' 
$B_d\to\pi^+\pi^-$, $B_s\to K^+K^-$ approach, which is not affected by
such problems. 

Hopefully, studies of the kind discussed in this paper will eventually 
guide us to the physics lying beyond the Standard Model.


\begin{thebibliography}{99}


\bibitem{ut}L.L. Chau and W.-Y. Keung, {\it Phys.\ Rev.\ Lett.}~{\bf 53}, 
1802 (1984); C. Jarlskog and R.~Stora, {\it Phys.\ Lett.}~{\bf B208}, 268
(1988). 
 
\bibitem{ckm}N. Cabibbo, {\it Phys.\ Rev.\ Lett.}~{\bf 10}, 531 (1963); 
M. Kobayashi and T.~Maskawa, {\it Progr.\ Theor.\ Phys.}~{\bf 49}, 652 (1973).
 
\bibitem{bisa}A.B. Carter and A.I. Sanda, {\it Phys.\ Rev.\ Lett.}~{\bf 45},
952 (1980); {\it Phys.\ Rev.}~{\bf D23}, 1567 (1981); I.I. Bigi and A.I. 
Sanda, {\it Nucl.\ Phys.}~{\bf B193}, 85 (1981).

\bibitem{revs}For reviews, see, for instance, {\it The BaBar Physics Book},
eds.\ P.F. Harrison and H.R.~Quinn (SLAC report 504, October 1998); Y. Nir, 
published in the Proceedings of the 18th International Symposium on 
Lepton--Photon Interactions (LP '97), Hamburg, Germany, 28 July--1 August 
1997, eds.\  A. De Roeck and A. Wagner (World Scientific, Singapore, 1998), 
p.\ 295 [hep-ph/9709301]; M. Gronau, {\it Nucl.\ Phys.\ Proc.\ 
Suppl.}~{\bf 65}, 245 (1998); R. Fleischer, {\it Int.\ J. Mod.\ 
Phys.}~{\bf A12}, 2459  (1997).

\bibitem{alph}See, for instance, M. Gronau, {\it Phys.\ Lett.}~{\bf B300},
163 (1993); J.P. Silva and L. Wolfenstein, {\it Phys.\ Rev.}~{\bf D49}, R1151
(1994); R. Aleksan {\it et al.}, {\it Phys.\ Lett.}~{\bf B356}, 95 (1995); 
F. DeJongh and P. Sphicas, {\it Phys.\ Rev.}~{\bf D53}, 4930 (1996); 
R. Fleischer and T. Mannel, {\it Phys.\ Lett.}~{\bf B397}, 269 (1997);
M. Ciuchini {\it et al.}, {\it Nucl.\ Phys.}~{\bf B501}, 271 (1997); 
P.S. Marrocchesi and N. Paver, {\it Int.\ J. Mod.\ Phys.}~{\bf A13}, 251 
(1998).

\bibitem{BsKK}R. Fleischer and I. Dunietz, {\it Phys.\ Rev.}~{\bf D55}, 259
(1997); C.S. Kim, D. London and T. Yoshikawa, {\it Phys.\ Rev.}~{\bf D57}, 
4010 (1998); R. Fleischer, {\it Phys.\ Rev.}~{\bf D58}, 093001 (1998).

\bibitem{FSI}See, for instance, L. Wolfenstein, {\it Phys.\ Rev.}~{\bf D52}, 
537 (1995); J.-M. G\'erard and J. Weyers, {\it Eur.\ Phys.\ J.}~{\bf C7},
1 (1999); M. Neubert, {\it Phys.\ Lett.}~{\bf B424}, 152 (1998); A.F. Falk, 
A.L. Kagan, Y. Nir and A.A. Petrov, {\it Phys.\ Rev.}~{\bf D57}, 4290 (1998);
D. Atwood and A. Soni, {\it Phys.\ Rev.}~{\bf D58}, 036005 (1998); 
A.N. Kamal, preprint ALBERTA-THY-01-99 (1999) [hep-ph/9901342].

\bibitem{BpiK}See, for instance, R. Fleischer and T. Mannel, 
{\it Phys.\ Rev.}~{\bf D57}, 2752 (1998); M.~Gronau and J.L. Rosner, 
{\it Phys.\ Rev.}~{\bf D57}, 6843 (1998); R. Fleischer, 
{\it Eur.\ Phys. J.}~{\bf C6}, 451 (1999); M. Neubert and J.L. Rosner,
{\it Phys.\ Rev.\ Lett.}~{\bf 81}, 5076 (1998); R. Fleischer and A.J.
Buras, preprint CERN-TH/98-319 (1998) [hep-ph/9810260].

\bibitem{Uspin-PAP1}R. Fleischer, preprint CERN-TH/99-78 (1999).

\bibitem{Bsclean}M. Gronau and D. London, {\it Phys.\ Lett.}~{\bf B253},
483 (1991); R. Aleksan, I. Dunietz and B. Kayser, {\it Z. Phys.}~{\bf
C54}, 653 (1992); R. Fleischer and I. Dunietz, {\it Phys.\ Lett.}~{\bf B387}, 
361 (1996).

\bibitem{DGamma}For a recent calculation of $\Delta\Gamma_s$, see 
M. Beneke, G. Buchalla, C. Greub, A. Lenz and U. Nierste, preprint 
CERN-TH/98-261 (1998) [hep-ph/9808385].

\bibitem{dunietz}I. Dunietz, {\it Phys.\ Rev.}~{\bf D52}, 3048 (1995).

\bibitem{wolf}L. Wolfenstein, {\it Phys.\ Rev.\ Lett.}~{\bf 51}, 1945 (1983).
 
\bibitem{blo}A.J. Buras, M.E. Lautenbacher and G. Ostermaier, {\it Phys.\
Rev.}~{\bf D50}, 3433 (1994).

\bibitem{nirsil}Y. Nir and D. Silverman, {\it Nucl.\ Phys.}~{\bf B345}, 
301 (1990).

\bibitem{bfm}A.J. Buras, R. Fleischer and T. Mannel, {\it Nucl.\ 
Phys.}~{\bf B533}, 3 (1998).

\bibitem{ddf1}A.S. Dighe, I. Dunietz and R. Fleischer, 
{\it Eur.\ Phys. J.}~{\bf C6}, 647 (1999).

\bibitem{ambig}See, for example, Y. Grossman and H.R. Quinn, 
{\it Phys.\ Rev.}~{\bf D56}, 7259 (1997); J.~Charles, A. Le Yaouanc, 
L. Oliver, O. P\`ene and J.-C. Raynal, {\it Phys.\ Lett.}~{\bf B425}, 375
(1998); A.S. Dighe, I. Dunietz and R. Fleischer, 
{\it Phys.\ Lett.}~{\bf B433}, 147 (1998).

\bibitem{charles}J. Charles, {\it Phys.\ Rev.}~{\bf D59}, 054007 (1999).

\bibitem{rf}R. Fleischer, in \cite{BpiK}.

\bibitem{BSW}M. Bauer, B. Stech and M. Wirbel, {\it Z. Phys.}~{\bf C29}, 
637 (1985) and {\bf C34}, 103 (1987).

\bibitem{ghlr}M. Gronau, O.F. Hern\'andez, D. London and J.L. Rosner,
{\it Phys.\ Rev.}~{\bf D52}, 6356 and 6374 (1995). 

\bibitem{groro}M. Gronau and J.L. Rosner, {\it Phys.\ Rev.}~{\bf D58},
113005 (1998). 

\bibitem{stone}S. Stone, talk given at WIN '99, Cape Town, 
South Africa, January 1999, to appear in the Proceedings.

\bibitem{BBL-rev}For a review, see G. Buchalla, A.J. Buras and M.E. 
Lautenbacher, {\it Rev.\ Mod.\ Phys.}~{\bf 68}, 1125 (1996).

\bibitem{bss}M. Bander, D. Silverman and A. Soni, {\it Phys.\ Rev.\
Lett.}~{\bf 43}, 242 (1979).

\bibitem{pen-calc}R. Fleischer, {\it Z. Phys.}~{\bf C58}, 483 
(1993) and {\bf C62}, 81 (1994); G. Kramer, W.F. Palmer and H. Simma, {\it Z. 
Phys.}~{\bf C66}, 429 (1995).

\bibitem{Buras-NLO}A.J. Buras, M. Jamin, M.E. Lautenbacher and P.H. Weisz,
{\it Nucl.\ Phys.}~{\bf B370}, 69 (1992); A.J. Buras, M. Jamin and M.E. 
Lautenbacher, {\it Nucl.\ Phys.}~{\bf B408}, 209 (1993).

\bibitem{georgi}H. Georgi, {\it Weak Interactions and
Modern Particle Theory} (Addison--Wesley Publishing Company, Redwood City,
California, 1984).

\end{thebibliography}
\end{document}